\documentclass[12pt]{article}
\usepackage{arxiv}
\usepackage[utf8]{inputenc} % allow utf-8 input
\usepackage[T1]{fontenc}    % use 8-bit T1 fonts
\usepackage[square,sort,numbers]{natbib}
\usepackage{varioref}
\usepackage[dvipsnames]{xcolor}
\usepackage[colorlinks]{hyperref}       % hyperlinks
\hypersetup{
linkcolor=BrickRed
,citecolor=Green
,filecolor=Mulberry
,urlcolor=NavyBlue
,menucolor=BrickRed
,runcolor=Mulberry
,linkbordercolor=BrickRed
,citebordercolor=Green
,filebordercolor=Mulberry
,urlbordercolor=NavyBlue
,menubordercolor=BrickRed
,runbordercolor=Mulberry
}%for the sake of ease for eyes
\usepackage[labelfont=bf]{caption}
\usepackage{subcaption}
\usepackage{enumitem}
\usepackage{url}            % simple URL typesetting
\usepackage{booktabs}       % professional-quality tables
\usepackage{amsfonts}       % blackboard math symbols
\usepackage{nicefrac}       % compact symbols for 1/2, etc.
\usepackage{microtype}      % microtypography
\usepackage{lipsum}		% Can be removed after putting your text content
\usepackage{graphicx}
\usepackage{natbib}
\usepackage{doi}
\usepackage{multirow}
\usepackage[bottom]{footmisc}
\usepackage{amsmath}
\usepackage{comment}
\newcommand\norm[1]{\left\lVert#1\right\rVert}
\newcommand{\angstrom}{\textup{\AA}}
\numberwithin{figure}{section}
\numberwithin{equation}{section}
\usepackage{amssymb}
\usepackage{tikz}
\usetikzlibrary{shapes.geometric,arrows}
\usetikzlibrary{calc}
\usetikzlibrary{positioning}
\usepackage[linesnumbered,ruled,vlined,algosection]{algorithm2e}
\usepackage{setspace}
\usepackage{indentfirst}
\tikzstyle{startstop} = [rectangle, rounded corners,  minimum height=1cm,text centered, draw=black]
\tikzstyle{io} = [trapezium, trapezium left angle=70, trapezium right angle=110, text centered, draw=black]
\tikzstyle{process} = [rectangle, minimum height=1cm, text centered, draw=black]
\tikzstyle{decision} = [diamond,  minimum height=1cm, aspect=2,text centered, draw=black]
\tikzstyle{arrow} = [thick,->,>=stealth]
\usepackage{amsmath}
\DeclareMathOperator*{\argmax}{arg\,max}

\title{Molecular modeling with machine-learned universal potential functions}

\date{\today}	% Here you can change the date presented in the paper title
\author{
	\textbf{Ke Liu \quad Zekun Ni \quad Zhengyu Zhou \quad Suocheng Tan \quad Xun Zou}\\ \textbf{Haoming Xing \quad Xiangyan Sun \quad Qi Han \quad Junqiu Wu \quad Jie Fan}\thanks{Corresponding author details: Accutar Biotechnology Inc., 760 Parkside Ave., Room 213, Brooklyn, NY 11226, USA} \\
	\\
	\textrm{Accutar Biotechnology} \\
	\texttt{jiefan@accutarbio.com}
}

\renewcommand{\headeright}{}
\renewcommand{\undertitle}{}
\hypersetup{
pdftitle={PP Paper},
pdfsubject={Machine Learning, Computational Chemistry, Computational Biology},
pdfauthor={AAA},
pdfkeywords={Machine learning, force field, molecular modeling, ligand docking, conformational search},
}

\begin{document}
\maketitle
\begin{abstract}
Molecular modeling is an important topic in drug discovery. Decades of research have led to the development of high quality scalable molecular force fields. In this paper, we show that neural networks can be used to train a universal approximator for energy potential functions. By incorporating a fully automated training process we have been able to train smooth, differentiable, and predictive potential functions on large-scale crystal structures. A variety of tests have also been performed to show the superiority and versatility of the machine-learned model.
\end{abstract}

% keywords can be removed

\section{Introduction} \label{secintro}

Molecular modeling started before the development of the modern digital computer, with one of the first simulations performed using wooden balls connected by springs in the basic ball-and-stick model ~\cite{molsimFrenkel}. With the development of computers that are one million times faster, the basic representation of molecules has not changed much in modern molecular modeling.

\begin{figure}
	\centering
	\begin{subfigure}[b]{0.6\textwidth}
		\centering
		\includegraphics[width=\textwidth]{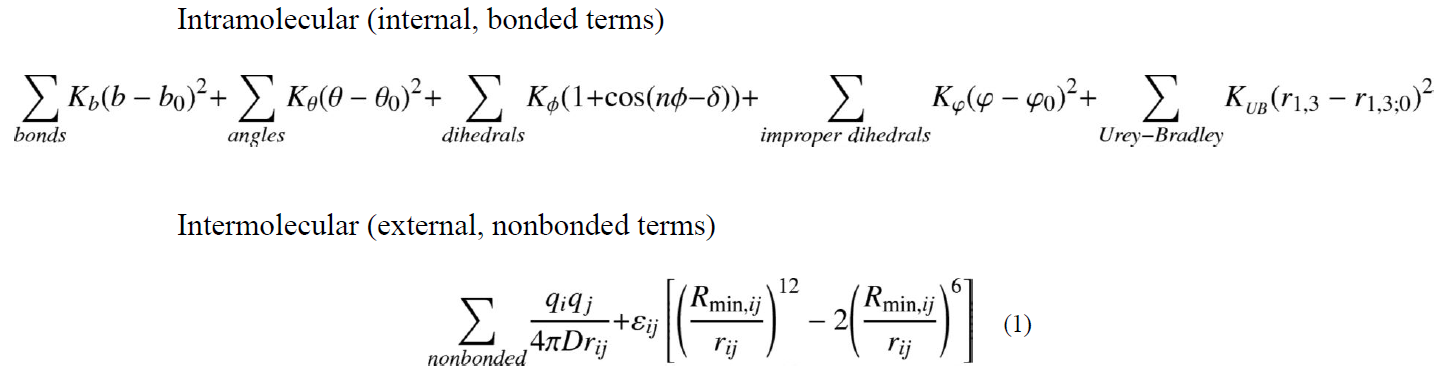}
		\caption{energy terms in the CHARMM force field~\cite{vanommeslaeghe2012automationcharmm}}
		\label{figcharmmenergy}
	\end{subfigure}
	\begin{subfigure}[b]{0.3\textwidth}
		\centering
		\includegraphics[width=\textwidth]{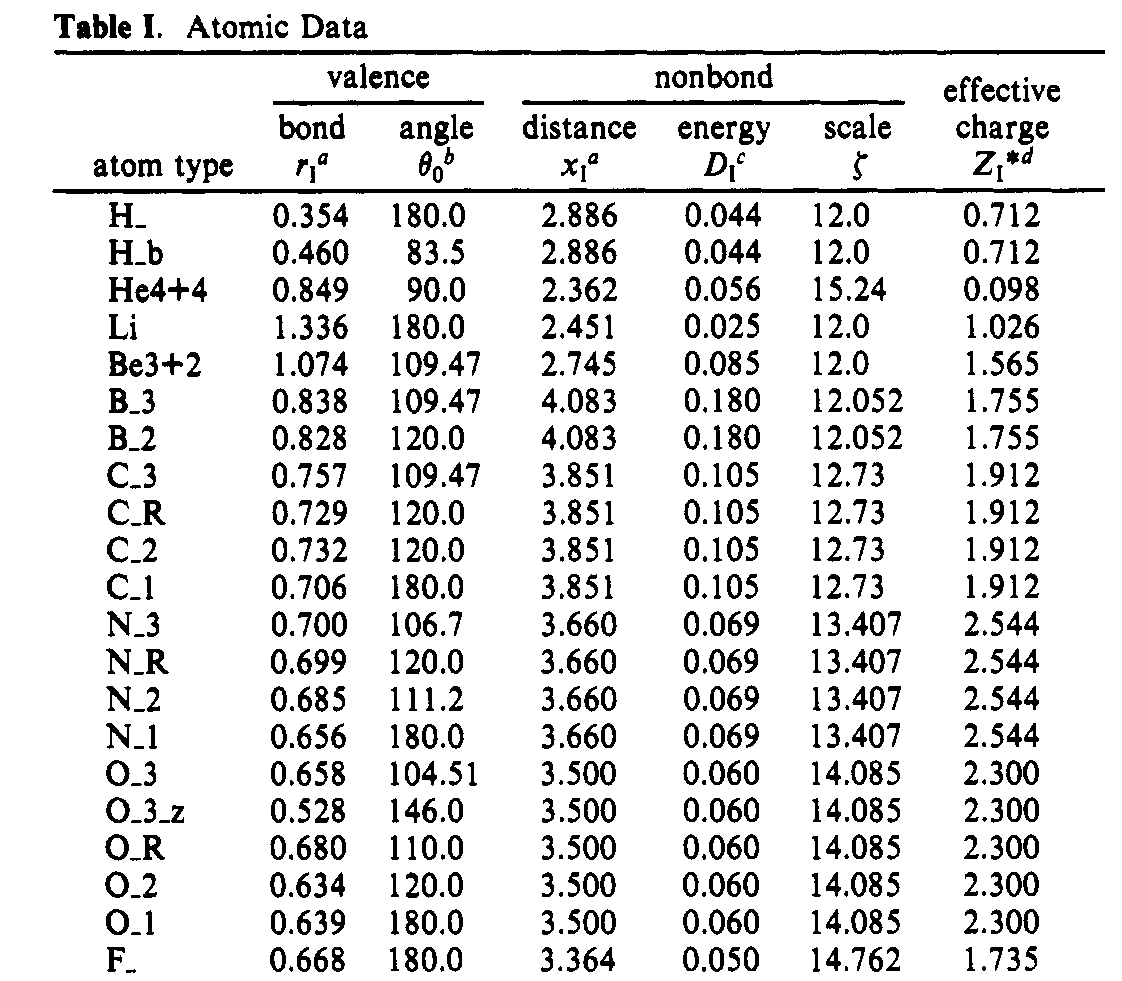}
		\caption{some atom types in the UFF force field~\cite{rappe1992uff}}
		\label{figuffatomtypes}
	\end{subfigure}
	\caption{Energy terms and atom typing in other force fields}
	\label{figcharmmuff}
\end{figure}

The basic way of molecular modeling is to define a \textit{force field} for quantizing the forces between molecular atoms. A force field usually covers the bond, angle, and dihedral tension forces of chemical compounds. Most modern force fields share the same structure and parametrization process. Energy terms are usually defined as a result of a compromise between physical intuition and computational feasibility. Atoms are categorized into tens of or sometimes hundreds of handpicked types. Parametrization is carried out by fitting the system to a specific dataset, usually higher-level computational results from a variety of molecules to ensure some desired properties such as co-planarity of aromatic rings, and plausible fit of bond length and angles. However, functions as complicated as atomic/molecular interactions may be beyond the expressiveness of quadratic and polynomial functions, or any other fixed combination of common mathematical functions. There are many existing force field implementations. \autoref{figcharmmuff} shows examples of energy terms and atom types of force field CHARMM General Force Field (CHARMM-CGENFF)~\cite{vanommeslaeghe2012automationcharmm} and the Universal Force Field (UFF)~\cite{rappe1992uff}.

The major drawback of existing force fields is that the atom types, energy term functions, and parameters are usually hand-fitted on small datasets. This makes the energy fields less generalized and usually only works well for the specifically tuned field. Correction terms usually need to be added to force fields when working with other systems or rare atom types, which complicates the system greatly. With the recent renaissance of neural networks and artificial intelligence came many attempts of using neural networks as a method of modeling molecular systems. Many of these attempts used neural networks to calculate a higher level of abstraction of molecules as a feature extraction tool to replace hand-picked fingerprints. The Graph Convolutional Network (GCN) is the preferred way to model molecular information as it allows a seamless and lossless way of representing the input molecule as a graph. ~\cite{kearnes2016molecular} and ~\cite{cheminet} use GCN to modeling the physical and pharmaceutical properties of molecular compounds. ~\cite{bresson2019two} uses GCN for molecule generation.

Schnet ~\cite{schutt2018schnet} is a machine learning framework for molecules, which encodes atom types using learned embeddings. The work showed such embeddings, when mapped by the first and second principal component, can be used to group atoms of the same group into the same cluster. This network was originally trained on the Quantum Machine 9 (QM9) dataset, which consists only of hydrogen, carbon, nitrogen, oxygen, and fluorine elements, but it is now trained on the Materials Project database that contains mainly inorganic compounds and 50,000 molecules. In the training process, each progression involves converting atom positions into absolute positions  $\boldsymbol{r}_i$ to describe the local chemistry environment of an atom. Some tricks such as calculating pairwise distances instead of using relative positions are employed to implement rotational invariance.

The benefit of the neural network-based approach is it can learn important features and interactions between atoms and bonds automatically. The main drawback is the lack of interpretability of the trained model, i.e., it is difficult to explain how the model works for a given input molecule. This makes the neural network a poor replacement for traditional force fields, as the latter have clear physical correspondence and are established on a variety of molecule modeling tasks.

In this paper, we propose a hybrid framework combining the strength of both the force field and neural network-based approaches. We define a neural network-based energy potential function which has the advantage of being trained on a large set of available crystal molecule structures, while keeping the benefits of traditional force fields as it is possible to do simulations that are molecular dynamics-like, and it can also be applied to side chain prediction and docking tasks. By having a trained neural network model, the model can generally adapt to all types of chemical systems such as protein-ligand complexes. Using a fully automatic training process, our method eliminates the manual tuning steps involved in traditional force fields. The dynamic negative sampling and bootstrapping algorithm used makes the potential functions have good local minima at ground truth conformations. As a result, the obtained neural network-based energy potential function shows superior performance on a variety of tasks over existing methods.

\section{Methodology} \label{secmethod}

\subsection{Atom Type Embedding} \label{subsecatomtype}
Atom types are used to distinguish the chemical identities of different atoms and group atoms with the same electric, chemical, and structural properties. Atom types are conventionally assigned using predefined rules that resemble corresponding chemical properties. A simple example of such rule is to define each (element, hybridization) pair as a separated atom type.

In this paper, we use an alternative known as learned atom type embedding. Instead of predefining rules of how to group or distinguish atoms, we train a neural network to generate embeddings of atom types. Such embeddings are then used in downstream models. In this way, the model can learn arbitrary complex atom type hierarchies, not limited to existing human chemical knowledge. The model also has the capability to encode more chemical information than just a single identification in such embedding. Different atom types may share some chemical properties, such as those with the same element number. Using an embedding-like distributed representation instead of hard-coded types could also leverage such similarities.

The atom embedder uses the graph representation of a molecule, where the vertices correspond to atoms and edges correspond to bonds. In addition, the following chemical features are extracted and associated with the vertices (atoms) and edges (bonds).

The vertex/atom features include:
\begin{itemize}
\item Element: the element type of the atom.
\item Charge: the electrostatic charge of the atom.
\item Radius: the van der Wells radius and covalent radius of the atom.
\item In ring: whether the atom is part of a ring.
\item In aromatic ring: whether the atom is part of an aromatic ring.
\end{itemize}

The edge/bond features include:
\begin{itemize}
\item Bond type: the type of the bond, one of $\{single, double, triple, aromatic\}$.
\item Same ring: whether the two atoms are in the same ring.
\end{itemize}

After an input molecule is transformed into a graph, the atom embedding of each atom is calculated by a graph convolution-like model~\cite{cheminet}. In each graph convolution layer, the embedding of each atom $E_i$ is updated by information from neighboring atoms:

\begin{equation}
  E_i^t = Reduce(\{ \{E_j^{t-1}, B(i,j)\} |j\in{Neighbor(i)}\})
\end{equation}

where the initial embeddings $E_i^0$ is the predefined atom features, $B(i, j)$ is the bond features. The $Reduce(\bullet)$ function is a set reduction as defined in section \label{subsubsecsymfuncapprox} for reducing information from a set of embeddings into one.

Each of the graph convolution aggregates the information for each atom from one bond further in the molecule graph. After $k$ steps the atom embeddings contain information from $k$ bonds away for each atom. The extracted atom embeddings are then fed into the downstream neural network, which is described in following sections.

\subsection{The Energy Potential model} \label{subsecenergypotentialmodel}

\subsubsection{Neural function approximator} \label{subsubsecneuralfuncapprox}

In this section we describe our neural network-based potential function approximator. The basic rationale is to train a smooth function approximation. This is a strong regularization term that prevents the model from overfitting and makes it suitable for gradient-based optimization.

The function approximator tries to learn a polynomial-like function for any input embedding. It consists of three layers:
\begin{enumerate}
\item The first layer feeds the input embedding through a conventional fully connected layer to allow a linear transformation on the input:
\begin{equation}
	X = Act(\boldsymbol{W}_0X_0 + \boldsymbol{b}_0)
\end{equation}
wherein $Act(x)$ is the activation function, $X_0$ is the input embedding, $\boldsymbol{W}$ and $\boldsymbol{b}$ are layer weights. We use a smooth activation function $Swish(x) = x\cdot sigmoid(x)$~\cite{ramachandran2017swish}.

\item The second layer first transforms the input to logarithmic scale, applying a scaling term, and then uses an exponential function to transform the output back. In this way the scaling term corresponds to the exponential factor of the polynomial:
\begin{equation}
	\boldsymbol{y}_i = exp(\boldsymbol{w}_{1i}\cdot log(1+X_i))
\end{equation}

\item The final layer applies a linear transformation of the polynomial output:
\begin{equation}
	\boldsymbol{z} = \boldsymbol{W}_2\boldsymbol{y}+\boldsymbol{b}_2
\end{equation}
\end{enumerate}

\subsubsection{Symmetrical function approximation} \label{subsubsecsymfuncapprox}

Most potential functions are symmetrical with respect to the exchange of input atoms. For example, for the van der Walls (vdW) potential function of an unbonded atom pair $i$ and $j$, the potential function $vdW(i,j)$ is symmetrical, i.e. $vdW(i,j)=vdW(j,i)$. However, for most neural networks, the input embeddings are ordered vectors, which violates the requirement of symmetry. To solve this problem, we apply the \textit{set reduction} function to those input groups whose orders are irrelevant.

Let $\boldsymbol{X={x_1,x_2,...,x_n}}$ be the set of input embeddings whose ordering is irrelevant. We first feed the embeddings through a fully connected layer to determine the importance of each embedding:
\begin{equation}
\boldsymbol{t} = \boldsymbol{Wx}+\boldsymbol{b}
\end{equation}

The importance weights are normalized via the Softmax function:
\begin{equation}
\boldsymbol{t}^{'}_{i} = \frac{e^{\boldsymbol{t}_i}}{\sum_{j}^{n} e^{\boldsymbol{t}_j}}
\end{equation}

Finally, all the embeddings are mixed according to calculated importance:
\begin{equation}
Reduce(X) = Reduce(\{X_1,X_2,...,X_n\})=\sum_{i}^{n} \boldsymbol{t}_i^{'}\boldsymbol{X_i}
\end{equation}

\subsubsection{Potential Terms} \label{subsubsecpotentialterm}
Theoretically, this function approximator model can be applied to any n-ary energy potential. In our experiments, we used the following potentials (\autoref{figpp}): 

\begin{figure}
	\centering
	\hfill
	\begin{subfigure}[b]{0.20\textwidth}
		\centering
		\includegraphics[width=\textwidth]{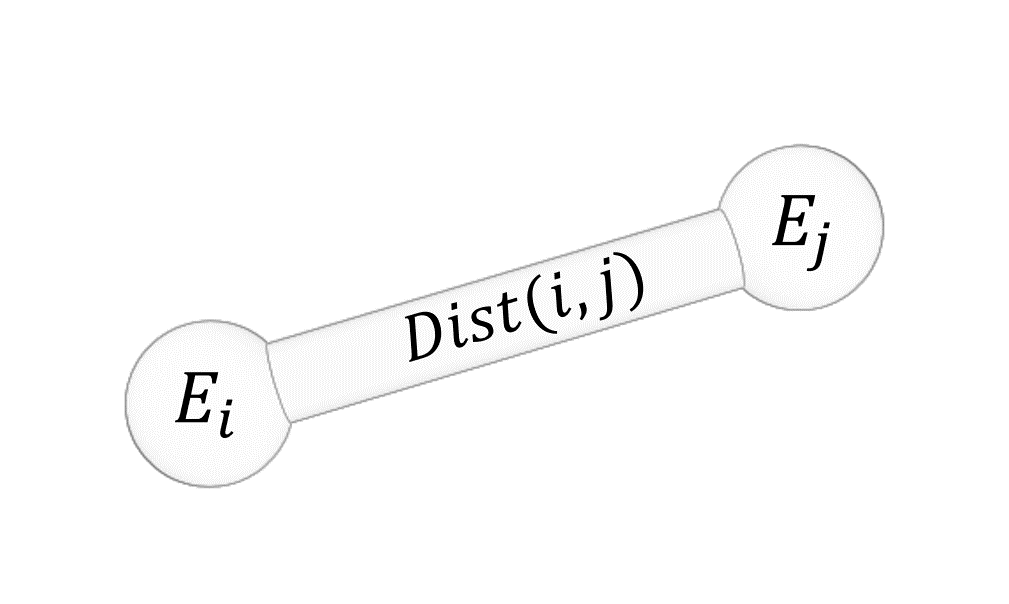}
		\caption{bonded}
		\label{figppbond}
	\end{subfigure}
	\hfill
	\begin{subfigure}[b]{0.20\textwidth}
		\centering
		\includegraphics[width=\textwidth]{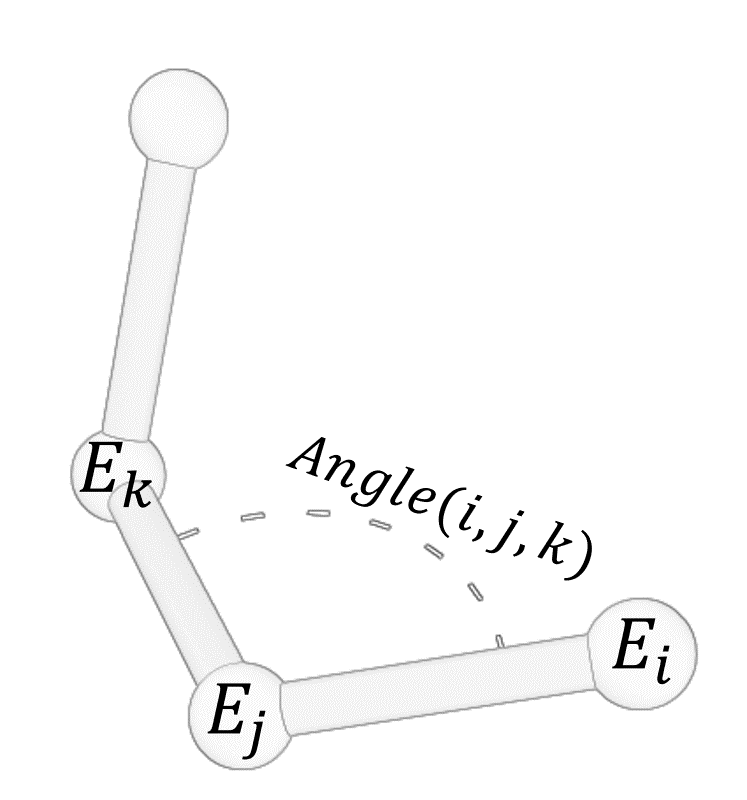}
		\caption{angle}
		\label{figppangle}
	\end{subfigure}
	\hfill
	\begin{subfigure}[b]{0.20\textwidth}
	\centering
	\includegraphics[width=\textwidth]{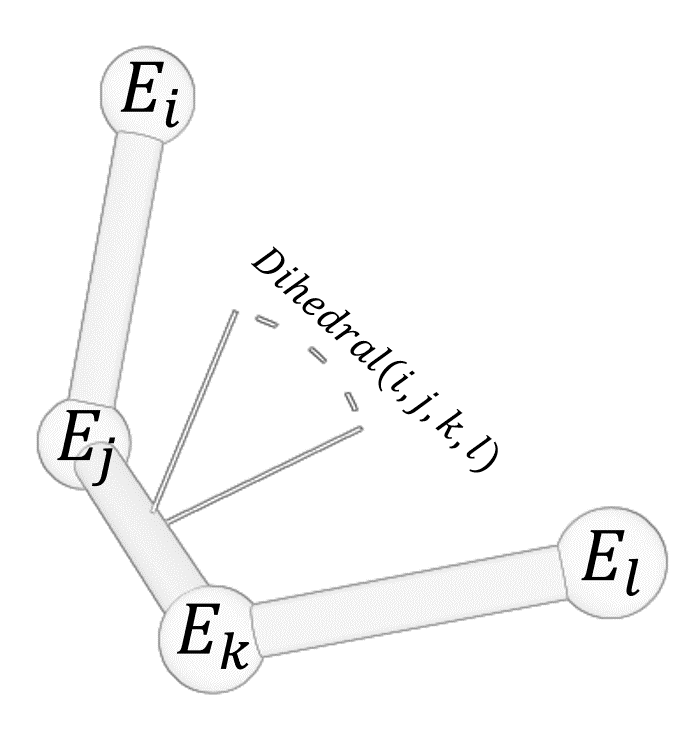}
	\caption{dihedral}
	\label{figppdiheral}
	\end{subfigure}
	\hfill

	\begin{subfigure}[b]{0.33\textwidth}
	\centering
	\includegraphics[width=\textwidth]{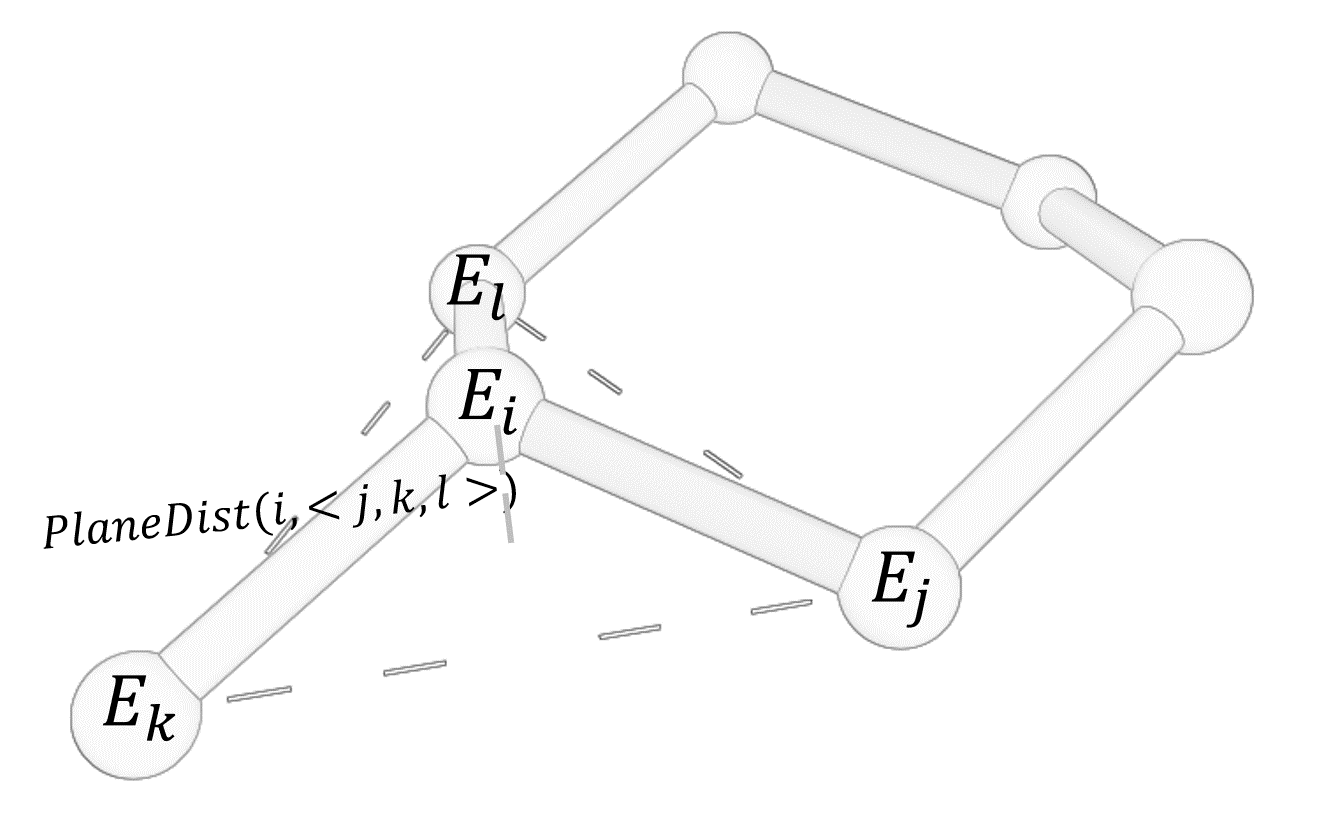}
	\caption{out-of-plane}
	\label{figppoop}
	\end{subfigure}
	\hfill
	\begin{subfigure}[b]{0.25\textwidth}
	\centering
	\includegraphics[width=\textwidth]{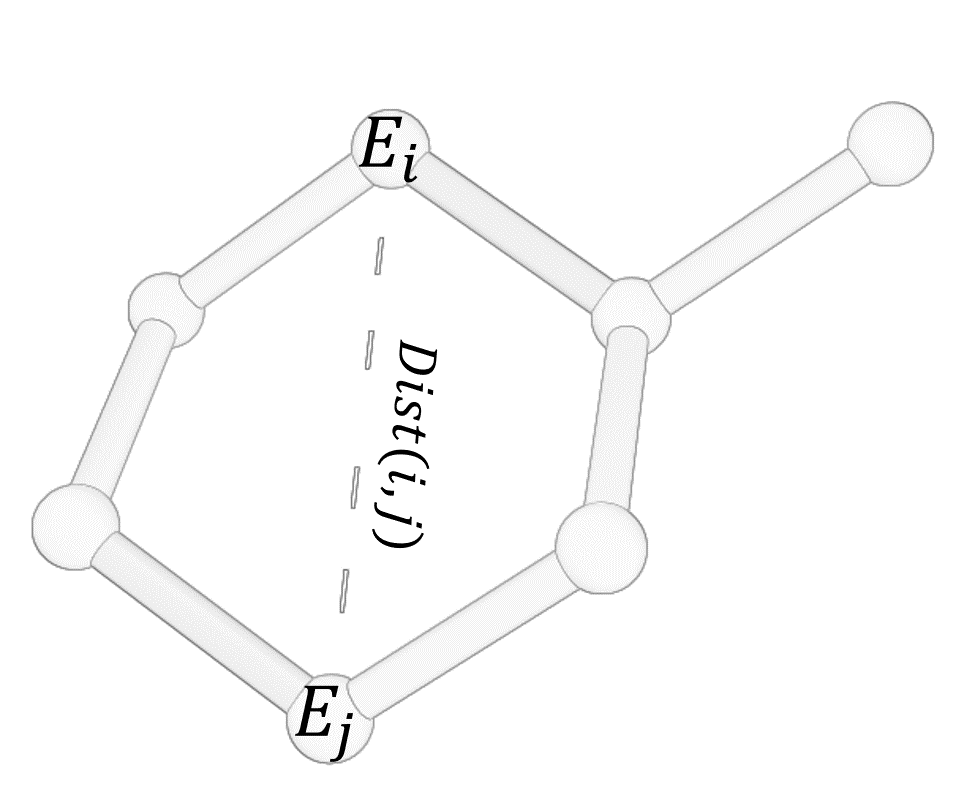}
	\caption{unbonded}
	\label{figppunbond}
	\end{subfigure}
	\hfill
	\begin{subfigure}[b]{0.15\textwidth}
	\centering
	\includegraphics[width=\textwidth]{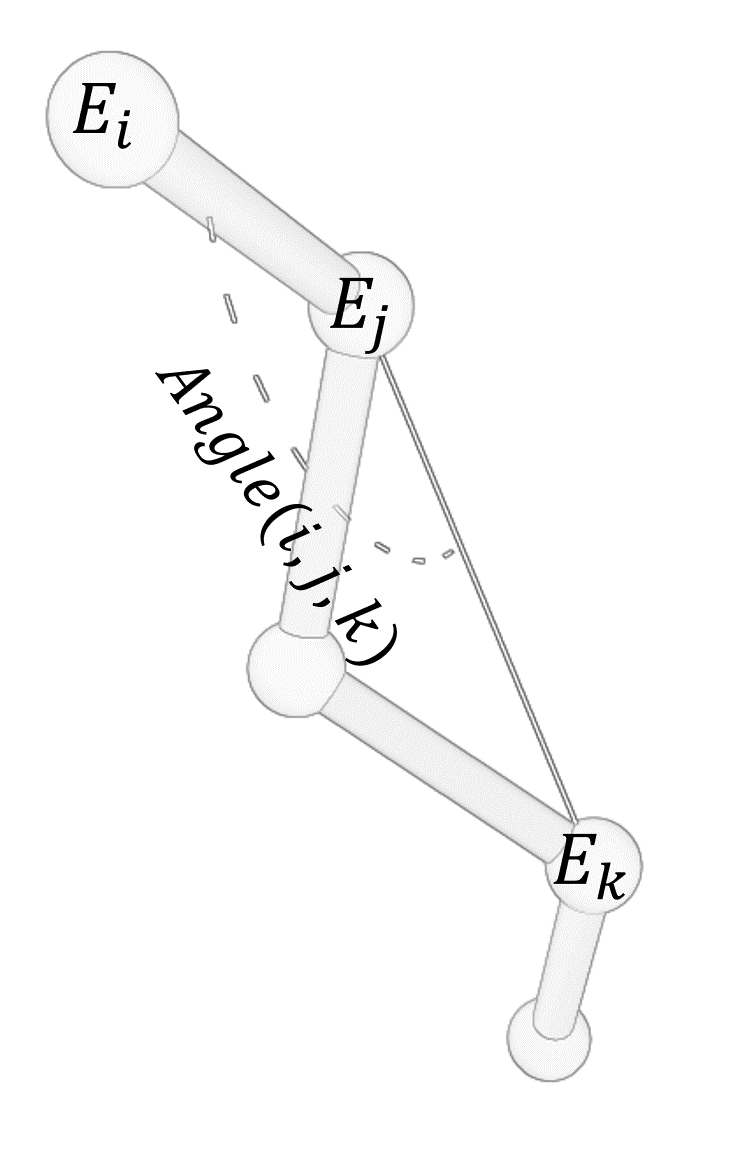}
	\caption{unbonded angle}
	\label{figppunangle}
	\end{subfigure}
	\hfill
	\begin{subfigure}[b]{0.25\textwidth}
	\centering
	\includegraphics[width=\textwidth]{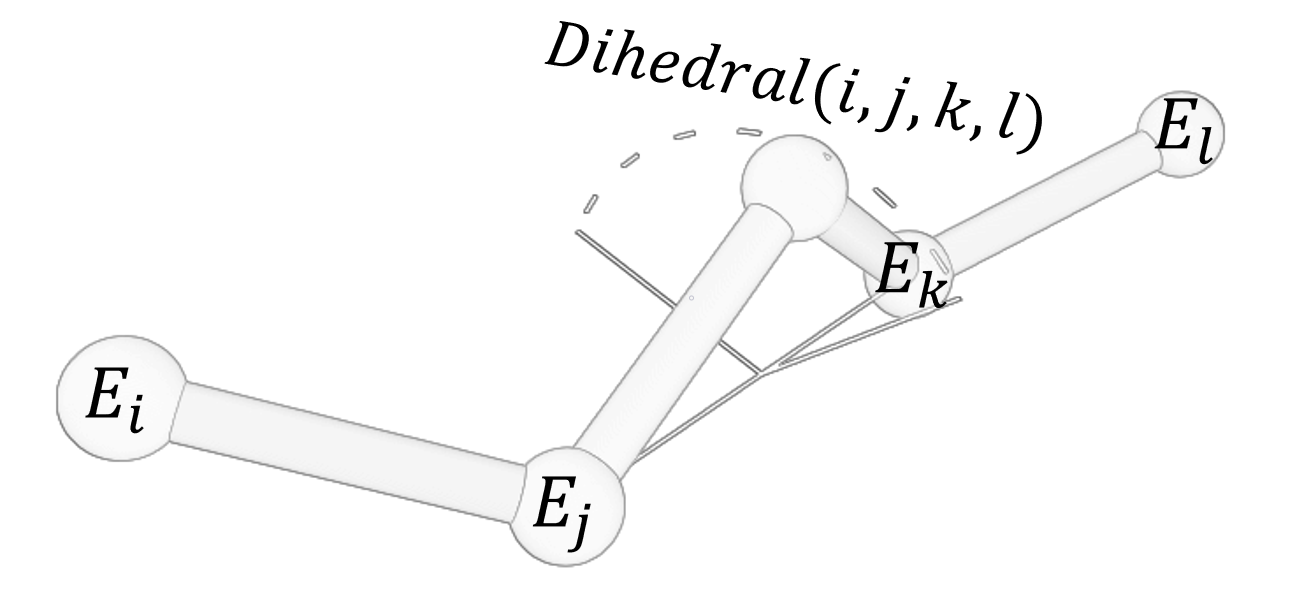}
	\caption{unbonded dihedral}
	\label{figppundihe}
	\end{subfigure}
	\hfill
	\caption{terms in the Potential Model}
	\label{figpp}
\end{figure}

\begin{itemize}
\item 	Bonded potential: this measures distance between pair of bonded atoms.
The bonded potential has the term for any bonded atoms $i-j$:
\begin{equation}
P_b (i,j)=F_b (Reduce(E_i,E_j ),Dist(i,j),I_b (i,j))
\end{equation}
wherein $E_i$ and $E_j$ are the atom embeddings of the two atoms, $Dist(i,j)$ is the Euclidean distance between the two atoms, $I_b (i,j)$ is the ideal bond length of the two atoms, and $F_b$ is the trained neural function approximator for the bond potential.

\item Angle potential: this measures bond tension between pair of bonds.
The angle potential has the term for any bonded atoms $i-j-k$:
\begin{equation}
P_a (i,j,k)=F_a (E_j,Reduce(E_i,E_k ),Angle(i,j,k),I_a (i,j,k))
\end{equation}
wherein $E_i, E_j$ and $E_k$ are the atom embeddings of the three atoms, $Angle(i,j,k)$ is the planar angle between the three atoms, $I_a (i,j,k)$ is the ideal bond angle of the three atoms, and $F_a$ is the trained neural function approximator for angle potential.

\item Dihedral potential: this measures dihedral angle tensions between two planes.
The dihedral potential has the term for any bonded atoms $i-j-k-l$:
\begin{equation}
P_d (i,j,k,l)=F_d (Reduce({E_i,E_j,E_k,E_l },{E_l,E_k,E_j,E_i }),Dihedral(i,j,k,l),I_d (i,j,k,l))
\end{equation}
wherein $E_i, E_j, E_k$ and $E_l$ are the atom embeddings of the four atoms, $Dihedral(i,j,k,l)$ is the dihedral between the two planes $<E_i,E_j,E_k>$  and $<E_j,E_k,E_l>$, $I_d (i,j,k,l)$ is the ideal dihedral angle of the four atoms, and $F_d$ is the trained neural function approximator for dihedral potential.

\item Out-of-plane potential: this measures the tension of planar atoms.
The out-of-plane potential has the term for atoms $j,k,l$ bonded to a central atom $i$:
\begin{equation}
P_{oop} (i,j,k,l)=F_{oop} (E_i,Reduce(E_j,E_k,E_l ),PlaneDist(i,<j,k,l>))
\end{equation}
wherein $E_i, E_j, E_k$ and $E_l$ are the atom embeddings of the four atoms, $PlaneDist(i,<j,k,l>)$ is the distance of the central atom $i$ to the plane $<j,k,l>$.
This term is added to atoms which is supposed to have planar bonds, such as $sp$ or $sp^2$ hybridized carbons.

\item Unbonded pairwise potential: this measures the distance between a pair of atoms without a connecting bond in between. This is similar to the bonded pairwise potential except it is for unbonded atoms. In other force fields, this term is usually divided into van Der Walls and electrostatic forces, which are then parameterized separately.
The unbonded pairwise potential has the term for any unbonded atoms pair $i$ and $j$:
\begin{equation}
P_{ub} (i,j)=F_{ub} (Reduce(E_i,E_j ),Dist(i,j))
\end{equation}
wherein $E_i$ and $E_j$ are the atom embeddings of the two atoms, $Dist(i,j)$ is the Euclidean distance between the two unbonded atoms.
	
\item Unbonded angle and unbonded dihedral: these terms are added to model the anisotropic electron distributions of polar atoms.They are similar to angle and dihedral potentials, except unbonded angle $P_{ua}$ is used for atoms $i,j,k$ wherein $i-j$ is bonded but $k$ is not bonded to $i,j$:
\begin{equation}
P_{ua} (i,j,k)=F_{ua} (E_i,E_j,E_k,Angle(i,j,k))
\end{equation}
unbonded dihedral $P_{ud}$ is used for atoms $i,j,k,l$ where $i-j$ and $k-l$ are two bonded pairs with no bonds in between:
\begin{equation}
P_{ud} (i,j,k,l)=F_{ud} (Reduce({E_i,E_j,E_k,E_l },{E_l,E_k,E_j,E_i }),Dihedral(i,j,k,l))
\end{equation}
\end{itemize}

\begin{figure}
\centering
\includegraphics[width=0.45\textwidth]{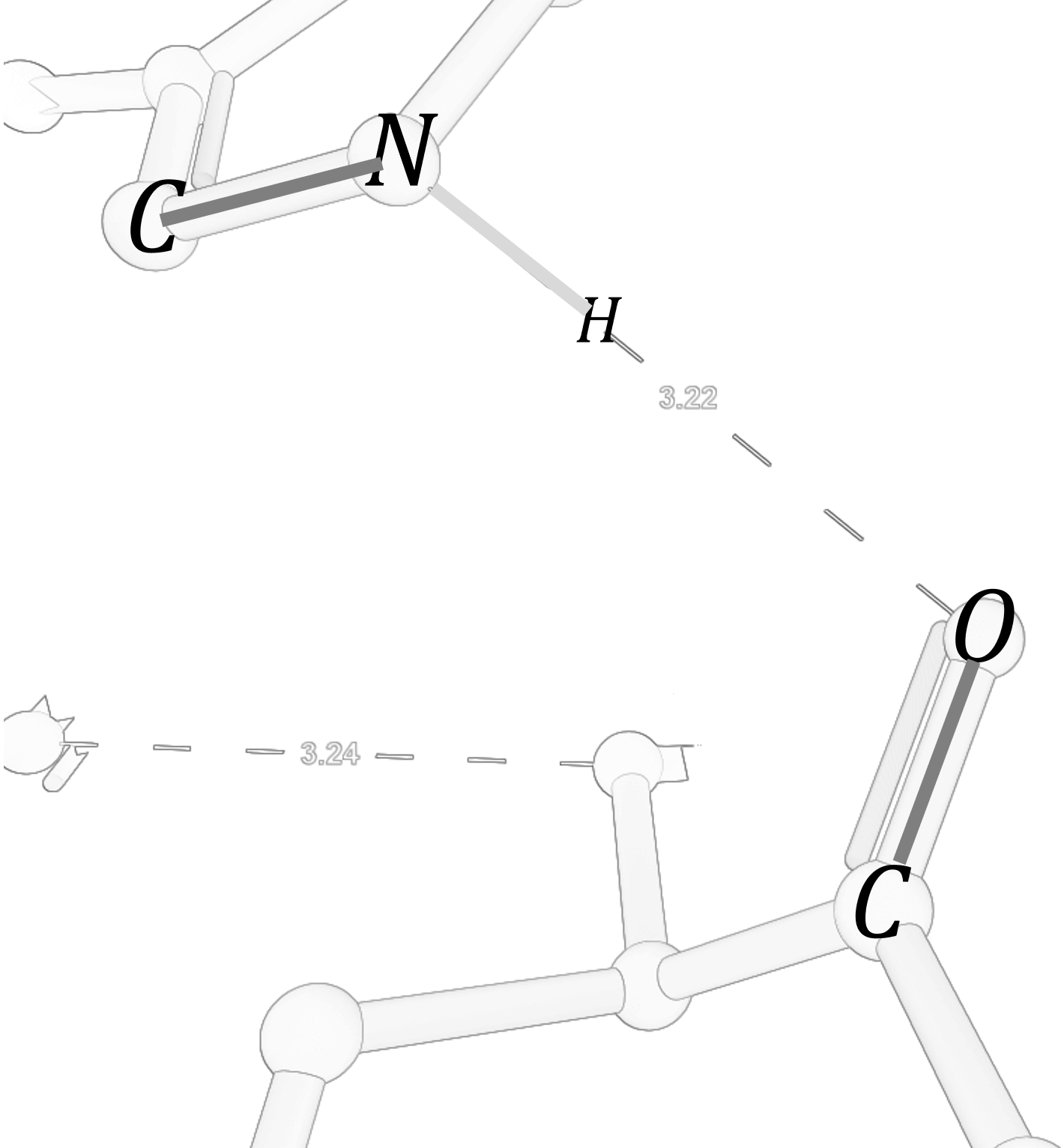}
\caption{Unbonded angle and dihedral in a hydrogen bond}
\label{figppshowcase}
\end{figure}

A typical situation wherein these two last terms are important is the hydrogen bond, which is very important in modeling intermolecular interactions such as ligand binding. In \autoref{figppshowcase}, notice that when hydrogen is not explicitly given, the hydrogen bond could only be deduced from the unbonded dihedrals, such as $Dihedral(C-N-O-C)$.  The anisotropy could be seen in the electron distribution of the oxygen atom in the carboxyl group. Thus, suppose we have an explicit H atom in the setting, then $Angle(H-O-C)$ being input enables the ability of the system to describe the polarizability of atoms and the formation criteria of a hydrogen bond. 

\subsubsection{Molecule potential}
The total potential function of a given molecule $m$ is the sum of all extracted potential functions of the molecule:
\begin{equation}
\label{ppsum}
\begin{split}
P(m;\boldsymbol{\theta})= &\sum_{(i,j)\in Bonds(m)} P_b (i,j) + \sum_{(i,j,k)\in Angles(m)} P_a (i,j,k) + \sum_{(i,j,k,l)\in Dihedrals(m)} P_d (i,j,k,l) + \\ &\sum_{(i,j,k,l)\in Planes(m)} P_{oop} (i,j,k,l) + \sum_{(i,j)\in Unbondeds(m)} P_{ub} (i,j) +\\& \sum_{(i,j,k)\in UnbondedAngles(m)} P_{ua} (i,j,k) +  \sum_{(i,j,k,l)\in UnbondedDihedrals(m)} P_{ud} (i,j,k,l)
\end{split}
\end{equation}
wherein $\boldsymbol{\theta}$ is the set of parameters of the function approximators. They are tuned during the training process and fixed after the training is done.

Given any molecule, such potential function can be defined. The information used to calculate the function is based on two parts of the molecule: one is the inherent graph structure of the molecule, such as elements of atoms, bonds, bond angles, and dihedrals. These are fixed regardless of the conformations of the molecule; the other one is the conformation of the molecule, which can be defined as the positions of atoms in the molecule. Hence, we can also write $P(m;\boldsymbol{\theta})=P(x,y,z;\boldsymbol{\theta})$ where $(x_i,y_j,z_i)$ is the three-dimensional(3D) coordinates of atom $i$.

A natural application of the potential function is to directly optimize the atom positions using gradient descent, just like most Molecular Dynamics(MD) methods wherein a Newtonian energy and force model is applied. It is an iterative process as described in the following algorithm:

\begin{algorithm}
%\setstretch{1}
\SetAlgoLined
\DontPrintSemicolon
\caption{Gradient descent for molecular conformation optimization}
\label{algomolconfopt}
\KwIn{A molecule $m$ given as coordinate vectors $(\boldsymbol{x},\boldsymbol{y},\boldsymbol{z})$}
\KwOut{One or several low-energy 3D conformation of the input molecule}
\SetKwFunction{GD}{GradientDescent}
\SetKwProg{Fn}{Function}{}{}
\BlankLine
\Fn{\GD{$m=(\boldsymbol{x},\boldsymbol{y},\boldsymbol{z})$}}{
	$(\boldsymbol{x}^{(0)},\boldsymbol{y}^{(0)},\boldsymbol{z}^{(0)}) \leftarrow (\boldsymbol{x},\boldsymbol{y},\boldsymbol{z})$\;
	$t \leftarrow 0$\;
	\While{stop condition not satisfied} {
		Calculate score $P(\boldsymbol{x}^{(t)},\boldsymbol{y}^{(t)},\boldsymbol{z}^{(t)};\boldsymbol{\theta})$ and corresponding gradient $\nabla P = [\frac{\partial P}{\partial \boldsymbol{x}^{(t)}},\frac{\partial P}{\partial \boldsymbol{y}^{(t)}},\frac{\partial P}{\partial \boldsymbol{z}^{(t)}}]$\;
		$\boldsymbol{x}^{(t+1)} \leftarrow \boldsymbol{x}^{(t)} - \alpha \frac{\partial P}{\partial \boldsymbol{x}^{(t)}}$ \;
		$\boldsymbol{y}^{(t+1)} \leftarrow \boldsymbol{y}^{(t)} - \alpha \frac{\partial P}{\partial \boldsymbol{y}^{(t)}}$ \;
		$\boldsymbol{z}^{(t+1)} \leftarrow \boldsymbol{z}^{(t)} - \alpha \frac{\partial P}{\partial \boldsymbol{z}^{(t)}}$ \;
		$t \leftarrow t + 1$\;
	}
	\Return{$(\boldsymbol{x}^{(t)},\boldsymbol{y}^{(t)},\boldsymbol{z}^{(t)})$}
}
\end{algorithm}

Notice that to make the gradient descent method possible, the function approximator becomes essential: It transfers the conventionally noncontinuous, non-smooth neural network models to a smooth, differentiable function. % TBD <- This really differentiate this work from others using trained atom features to calculate properties: By contrast, some works (OrbNet …) could only reproduce single-point energy on computationally generated molecular conformations with higher level of quantum chemical theory. TODO: Plot some potential function terms <- Maybe compare it with some conventional function forms. (also, we can compare it to some auto-parametrization method that takes QM calculations as input)

\subsection{Negative sampling}
The training process of the potential function $P(m;\boldsymbol{\theta})$ involves determining the parameters $\boldsymbol{\theta}$. It is relatively easy to obtain a large set of ground truth conformations, e.g. from crystal structures of ligands and proteins. Negative samples can be easily generated, by distorting existing conformations. However, most of such randomly generated distortions will lead to trivial negative samples, such as those with clashing (overlapping) atoms. Such examples contribute little value to model training.

To resolve this problem, we use different strategies to sample negative examples. Such samples, by design, are intended to be at some good local minima of the conformation space. The strategies we used for the model training are described below:
\begin{itemize}
	\item Self-iterative training: this is a general way to make the potential approximators converge to a point where ground truth molecules have the best potential. For any given molecule $m$, we use the gradient descent algorithm to optimize the conformation and get molecule $m'$. If the positional deviation $\lvert m'-m \rvert $ is larger than a predefined threshold, it means the model converges to a wrong local minimum. We then add $m'$ to our negative training examples.
	\item Distorted side chain conformations: we use a rotamer library~\cite{sidechainpaper} to sample different rotamers of side chain conformations. For those rotamers far from the ground truth position, we add them to the negative training examples. We also apply gradient descent to those rotamers and add the results which are still far from the ground truth conformation to the negative examples. 
	\item Distorted backbone conformations: we distort the backbone atoms randomly with backbone libraries. Gradient-descent-optimized examples are also added.
	\item Docked conformations: Firstly, we use our docking algorithm to dock ligands to protein pockets. Then we add docked conformations that have a large difference from the crystal structure to the negative training examples.
\end{itemize}

The negative sampling process involves using the model we currently have, and different sampling algorithms as described in \autoref{secsidechain} and \autoref{secdocking}. This creates a self-dependency (and a genetic iteration or evolution) of the model. To make this process possible we train the model in a bootstrapping setting. Formally, the algorithm is: 

\begin{algorithm}
\SetAlgoLined
\DontPrintSemicolon
\caption{Negative sampling and modeling training}
\label{algomodeltraining}
\KwIn{A set of ground truth conformations of molecules and proteins $\boldsymbol{P}$, iteration count $I$.}
\KwOut{A predictive model in the form of a potential function $P(m;\boldsymbol{\theta})$}
\SetKwFunction{Random}{Random}
\SetKwFunction{NegativeSample}{NegativeSample}
\BlankLine
$\boldsymbol{\theta} \leftarrow \Random{}$\;
generate the initial set of negative samples $\boldsymbol{N_0}$\ \tcp*{this does not require a model}
$i \leftarrow 0$\;
\While{$i < I$}{
	train model $\boldsymbol{\theta_i}$ using dataset $\boldsymbol{(P,\{N_0, N_1, ..., N_i\})}$\;
	$\boldsymbol{N_{i+1}} \leftarrow \NegativeSample{$\boldsymbol{\theta_i}, P$}$
	$i \leftarrow i + 1$\;
}
\Return{$\boldsymbol{\theta}_I$}
\end{algorithm}

\subsection{Loss functions}
There are primarily two classes of loss functions used in the training of potential function approximators:
\begin{itemize}
	\item Ranking loss: for each pair of molecule conformations $(m_a,m_b)$ wherein it is known (from ground truth) that conformation $m_a$ is more energetically stable than $m_b$, our goal is to ensure that our potential function has the relation $P(m_a;\boldsymbol{\theta)}<P(m_b;\boldsymbol{\theta)}$. Hence the loss function is defined as $ReLU(P(m_b;\boldsymbol{\theta)}-P(m_a;\boldsymbol{\theta)})$.
	\item Gradient loss: for conformations for which we have good confidence in their stability, such as conformations in crystal structures, we want our potential function to converge to this conformation, at least \textit{locally}. One way to achieve this is to make the gradient of the potential function at the targeted conformation approach zero. Hence the loss function is defined as a squared loss over the gradient of the targeted atom positions, i.e.
\begin{equation}
L=(\frac{\partial P(\boldsymbol{m;\theta})}{\partial \boldsymbol{m}})^2=(\partial \frac{P(\boldsymbol{x,y,z;\theta})}{\partial X}+\frac{\partial P(\boldsymbol{x,y,z;\theta})}{\partial \boldsymbol{y}}+\frac{\partial P(\boldsymbol{x,y,z;\theta})}{\partial \boldsymbol{z}})^2
\end{equation}
\end{itemize}

\section{Application}
\subsection{Molecule conformation optimization}\label{molconfopt}
Once the potential models are defined, any molecule conformation can be directly optimized, as the functional approximator is designed to be smoothly differentiable. This process resembles molecular dynamics simulations. 
%TODO: Exp: deviation between optimized conformations and crystal structures.

This gradient descent scheme is efficient in finding local minima of conformational energy. However, as in Molecular Dynamics simulation, the problem is the non-convexity and the existence of numerous such local minima, making it hard to use gradient descent to cross high-energy barriers between distant conformations. For other tasks like sidechain conformation prediction, merely getting a set of independent local minima of sidechains is insufficient for the task. We then use an iterative sampling-then-optimization strategy as a general idea to circumvent this issue.

For any input molecule $m$ and differentiable potential function $p(m)$. We optimize the molecule using a general algorithm described as \autoref{algogenconfopt}.

\begin{algorithm}
\SetAlgoLined
\DontPrintSemicolon
\caption{General conformation optimization}
\label{algogenconfopt}
\KwIn{input molecule $m$, target diffrentiable potential function $p(m)$}
\KwOut{optimized conformation of $m$}
\BlankLine
\While{not converged}{
sample initial molecule conformations of $m$: $m_1,m_2,...,m_n\in M$ diversely in molecular conformation space.\;
\ForEach{$m_i \in \{m_1,m_2,...,m_n\}$}{perform gradient descent optimization using local gradient $\frac{\partial p(m_i)}{\partial m_i}$}
sample and modify parts of conformations $m_1,m_2,...,m_n$ using a discrete sampling method\;
}
\end{algorithm}

Since the potential function term $p(m)$ is additive in terms of atom embeddings, it can be summed by either the whole molecule or some parts of a molecule. This makes the sampling algorithm much more flexible. In practice, we may start with sampling only part of a molecule and then extend the sampled part, as integrated in algorithms in the following sections.

\subsection{Sidechain conformation prediction}\label{secsidechain}

The side chain conformation problem seeks to predict side chains of all amino acids given their fixed backbone conformations. This is a good testbed for molecular dynamics models in protein context. We examine this problem in the setting of leave-one-out prediction. That is to predict the side chain of a single amino acid with the environment fixed.

To effectively sample side chain conformations, we first build a backbone-independent rotamer library of side chain conformations~\cite{sidechainpaper}, by which we reduce the leave-one-out prediction problem into two stages. The first stage is to test all existing rotamers of the side chain of an amino acid; the second stage is then to fine-tune the best rotamers in the first stage, as shown by the algorithm below:

\begin{algorithm}
\SetAlgoLined
\DontPrintSemicolon
\caption{Leave-one-out side chain prediction (LOO prediction)}
\label{algoloopredict}
\KwIn{one amino acid in a protein $a$}
\KwOut{a set of stable conformations of the amino acid, sorted by their predicted energy potential}
\SetKwFunction{Perturb}{Perturb}
\SetKwFunction{GD}{GradientDescent}
\SetKwFunction{sort}{Sort}
\SetKwFunction{top}{Top}
\BlankLine
$\boldsymbol{M} \leftarrow Rotamers(a)$\; \label{algolinerotamer}
\While{stop criterion not satisfied}{
	$\boldsymbol{M}' \leftarrow \emptyset$; \;
	\ForEach{$m \in \boldsymbol{M}$}{
		$m' \leftarrow \Perturb{$m$}$\; \label{algolineperturb}
		$\boldsymbol{M}' \leftarrow \boldsymbol{M}' \cup \{ m' \}$\;
	}
	$\boldsymbol{M}' \leftarrow \sort{$\boldsymbol{M}'$}$ \tcp*{Sort by potential function $p(m)$}
	$\boldsymbol{M}' \leftarrow \top{$\boldsymbol{M}', K$}$\;
	$\boldsymbol{M}'' \leftarrow \emptyset$; \;
	\ForEach{$m \in \boldsymbol {\boldsymbol{M}'}$}{
		$m' \leftarrow \GD{$m$}$\;
		$\boldsymbol{M}'' \leftarrow \boldsymbol{M}'' \cup \{ m' \}$\;
	}
	$\boldsymbol{M} \leftarrow \boldsymbol{M}''$\;
}
\Return{$\boldsymbol{M}$}
\end{algorithm}

The rotamer library in \autoref{algolinerotamer} is a small diverse set of potential conformations of each amino acid, generated from the training dataset as described in ~\cite{sidechainpaper}. The perturbation step in \autoref{algolineperturb} randomly perturbates the dihedral angles of the amino side chain in small steps. This is used to cross the barrier of many different dihedral configurations.

\subsection{Ligand conformation generation} \label{secligandgen}

The ligand conformation generation problem seeks to generate correct 3D conformations of a given ligand structure depiction. It is usually used as a preparation step for downstream applications such as docking and property prediction. We solve the problem by taking the advantage of potential functions as different parts of the molecule can be independently sampled and having their potential score summed.

\begin{figure}
	\centering
	\hfill
	\begin{subfigure}[b]{0.47\textwidth}
		\centering
		\includegraphics[width=\textwidth]{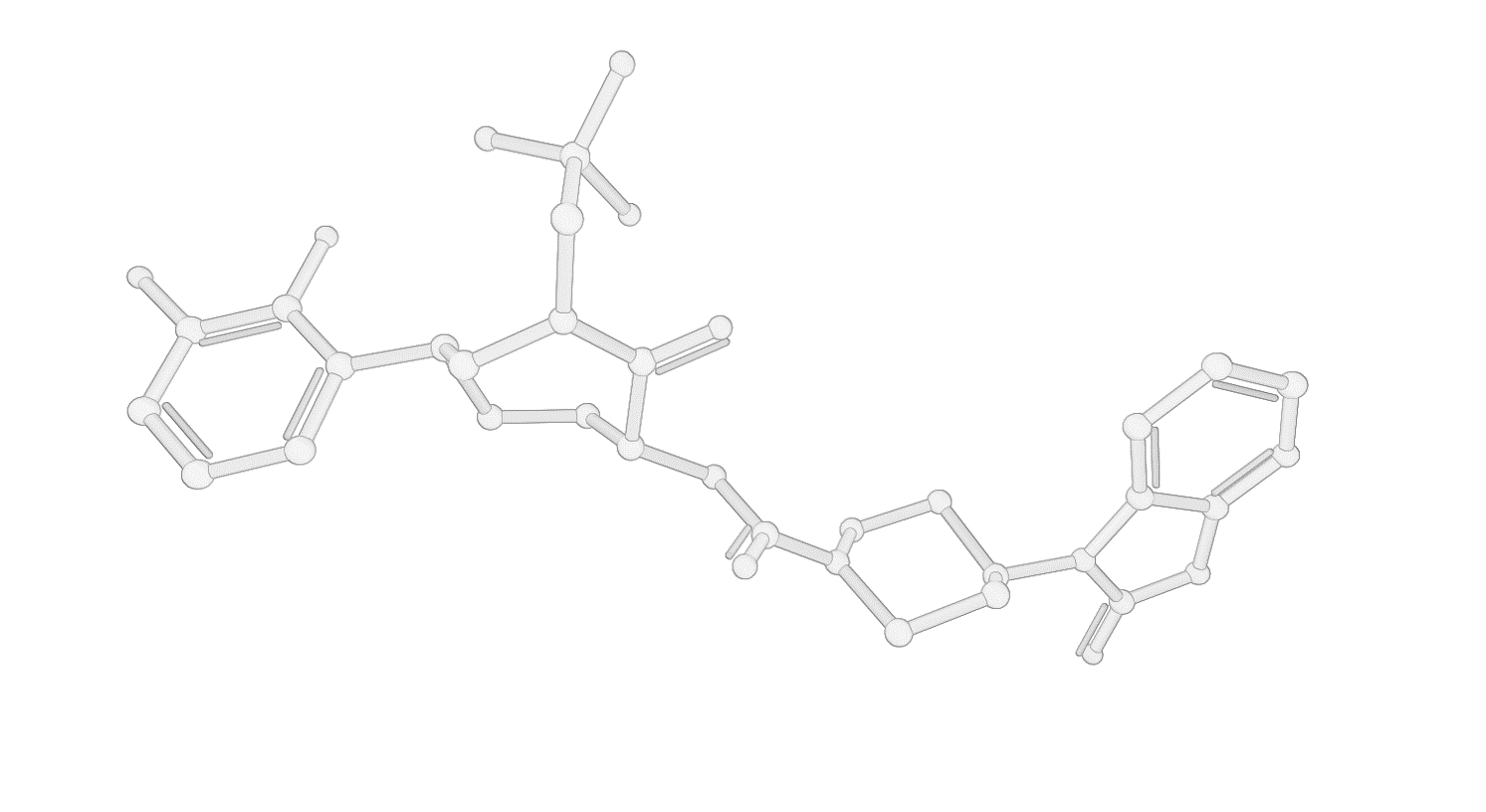}
		\caption{Ligand (N7R in PDB 3N7R)}
	\end{subfigure}
	\hfill
	\begin{subfigure}[b]{0.47\textwidth}
		\centering
		\includegraphics[width=\textwidth]{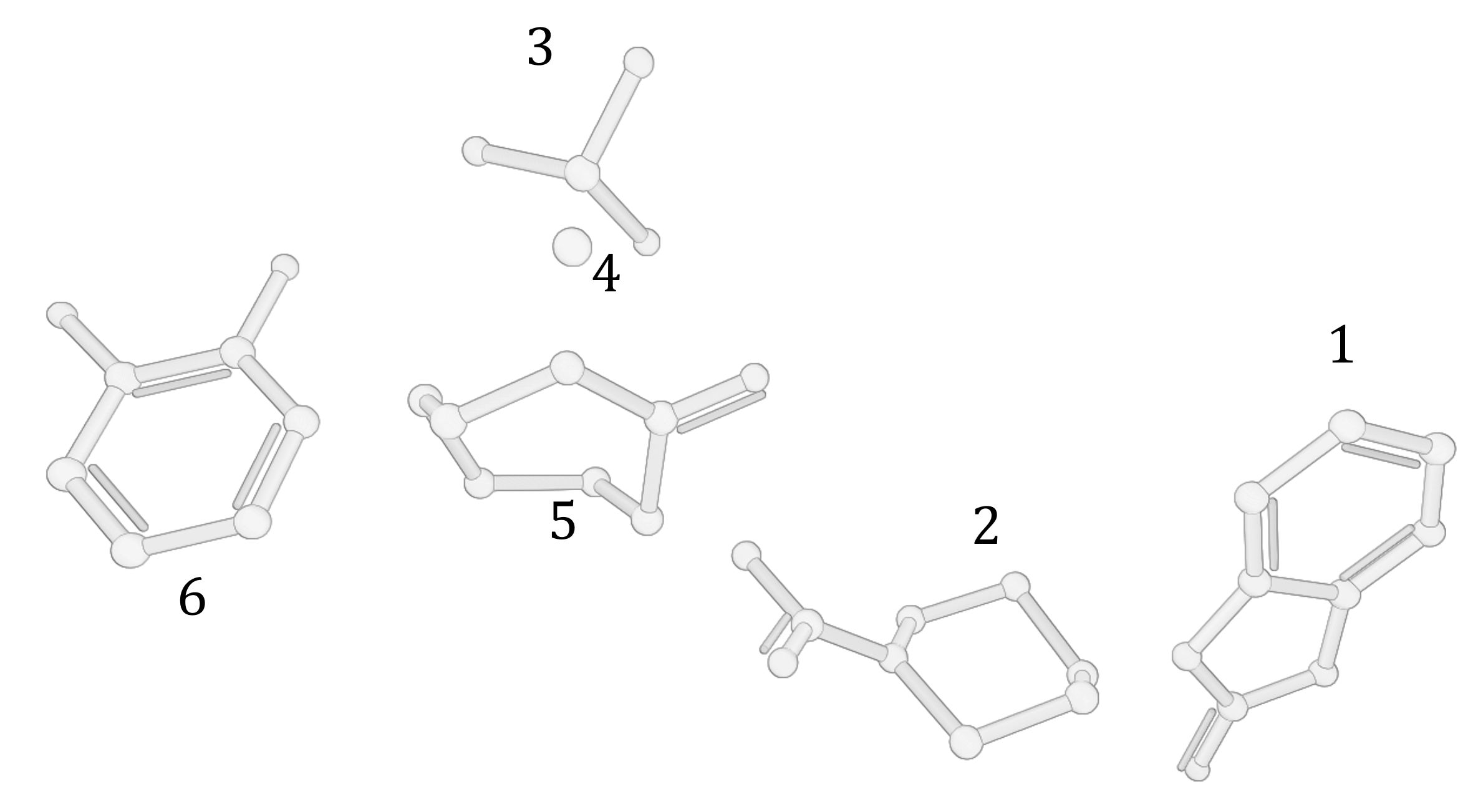}
		\caption{Rigid components of N7R}
	\end{subfigure}
	\hfill
	\caption{Ligand preparation (division)}
	\label{figligprep}
\end{figure}

First, the input compound is divided into rigid components connected by rotatable bonds, as illustrated in \autoref{figligprep}. Each component is independently generated by repetitively sample atom conformations near to currently sampled atoms, starting from an empty conformation set. We maintain a set of partial conformations $\boldsymbol{M}$ which has the same set of sampled atoms. We repetitively decide the next atoms to sample and extend the partial conformations until the partial set equals the full set of the atoms of the input molecule. The formal procedure is described in \autoref{algoligprep}.

\begin{algorithm}
\SetAlgoLined
\DontPrintSemicolon
\caption{Ligand conformation generation for rigid components}
\label{algoligprep}
\KwIn{Graph of the molecule structure, divided into rigid components $\boldsymbol{C}=\{c_i\}$}
\KwOut{Generated set of conformations for each rigid component}
\SetKwFunction{CompSize}{CompSize}
\SetKwFunction{FES}{FindExpandSet}
\SetKwFunction{Choose}{RandomSelect}
\SetKwFunction{FMain}{LigandGen}
\SetKwFunction{PlaceAtom}{PlaceAtom}
\SetKwFunction{PlaceRing}{PlaceRing}
\SetKwFunction{GD}{GradientDescent}
\SetKwProg{Fn}{Function}{}{}
\BlankLine
\Fn{\FES{$A, G(V,E)$}}{
	\uIf{$\exists$ Atom $a \notin A$ and $\exists$ $a' \in A$ where $(a, a') \in E$ and the position of $a$ could be determined by its hybridization configuration and existing atoms in $A$}{\Return $a$}
	\uElseIf{$\exists$ Ring $R$ s.t. $\ge 3$ atoms $\in R$ are in $A$}{\Return R}
	\Else{
		$A' \leftarrow \argmax_{b}{\CompSize{b},\forall a\in V\setminus A,b \in A,(a,b) \in E}$\;
		\tcp*{Find the atom $a\in V\setminus A$ which has most bonded atoms in $A$}
		\Return{\Choose{$A'$}} \tcp*{choose randomly from ties}
	}
}
\BlankLine
\Fn{\FMain{$\boldsymbol{C}$}}{
\ForEach{$c = G(V,E) \in \boldsymbol{C}$}{
	$A \leftarrow \Choose{$V$}$\;
	$\boldsymbol{M} \leftarrow \{(0,0,0)\}$\;
	\While{ $A \neq V$}{
		$X \leftarrow \FES{$A, G(V,E)$}$\;
		\ForEach{$m \in \boldsymbol{M}$}{
			\Switch{typeof ($X$)}{
				\uCase{Atom}{
					$m'=\PlaceAtom{$m, X$}$
					\tcp*{Place the bonded atoms of $X$ according to its hybridization (e.g. triangular for $sp^2$, linear for $sp$)}
				}
				\uCase{Ring}{
					$m'=\PlaceRing{$m, X$}$
					\tcp*{calculate the average plane of sampled atoms in the ring, then calculate the circumcircle of sampled atoms of their projection on the average plane, place remaining unsampled atoms of the ring on the equally divided points on the circumcircle}
				}
			}
			$m'' \leftarrow \GD{$m'$}$\;
			$\boldsymbol{M}' \leftarrow \boldsymbol{M}' \cup \{ m'' \}$\;
		}
		$\boldsymbol{M} \leftarrow \boldsymbol{M}'$\;
	}
}
\Return{$\boldsymbol{M}$}
}
\end{algorithm}

After the conformations are generated for each rigid component, a clash-avoiding sampling algorithm combined with gradient descent is used to connect the rigid components and sample good dihedral angle configurations of the rotatable bonds.

\subsection{Ligand Docking} \label{secdocking}

Molecular docking refers to the problem of computing the correct conformation of a ligand in a specified region of a protein (known as a pocket). We use the anchor-and-grow method for molecular docking. The input ligand is divided into rigid components connected by rotatable bonds (\autoref{figligprep}), similar to the first step in \autoref{secligandgen}. Then we repetitively place rigid components into the docked conformations, connecting them with existing atoms by rotating the dihedral of the connecting bond. During the docking process, only the position of parts of the final molecule is determined for each candidate conformation. The sampling algorithm repetitively extends docked components one by one to obtain the final docking result.

\begin{figure}
	\centering
	\includegraphics[width=\textwidth]{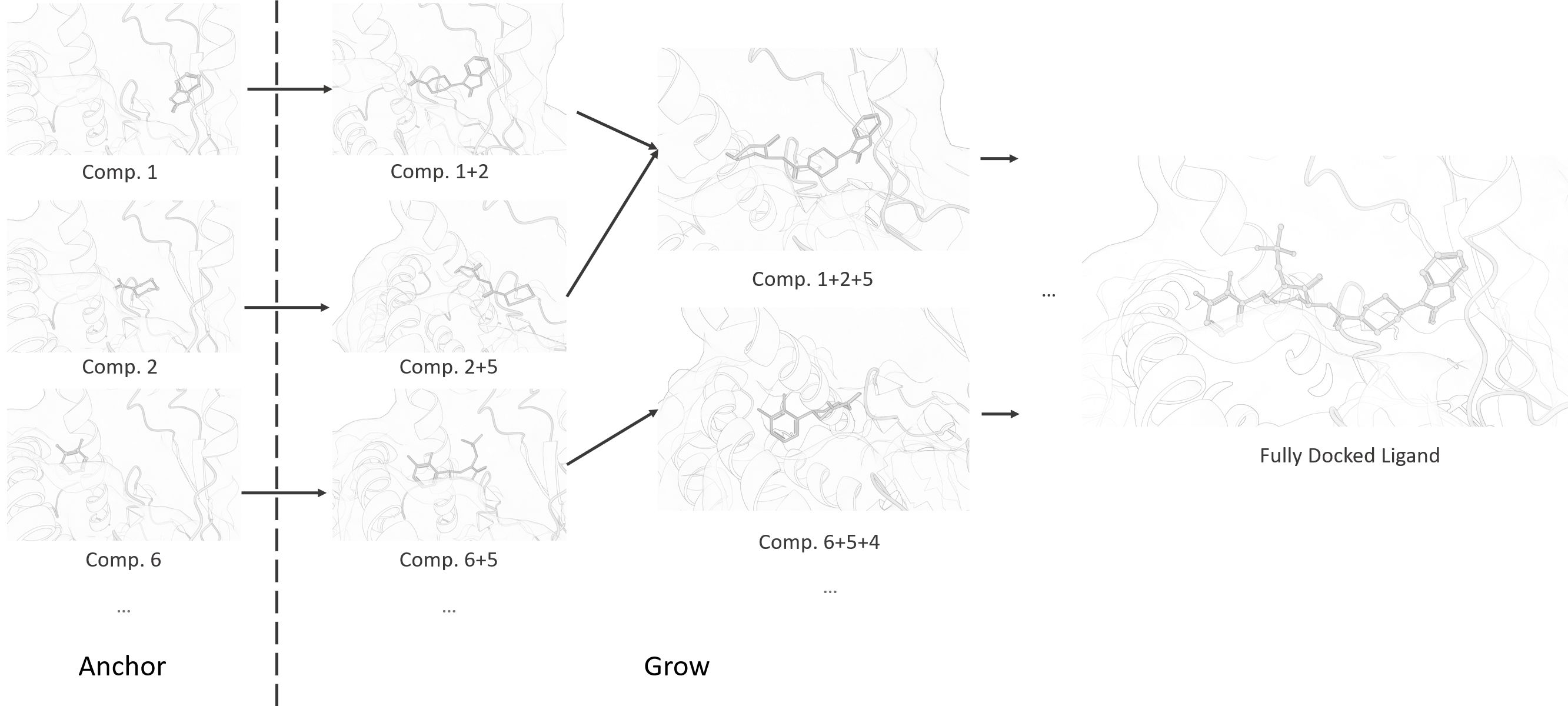}
	\caption{Anchor and grow algorithm for ligand docking (illustrative)}
	\label{figanchgrow}
\end{figure}

The docking algorithm is divided into the anchor and grow phases. In the anchor phase, the algorithm finds the best docking positions of each of the ligand's rigid components, resembling a simple rigid docking algorithm. In the grow phase, the algorithm extends existing partial docking conformations with other rigid components and optimizes dihedral angles in this process. Two compatible conformations may also be merged. The grow phase is organized in a beam search setting. This process is illustrated in \autoref{figanchgrow}.

The formal algorithm is shown in \autoref{algoanchorgrow} and \autoref{algodock}. The input of the docking algorithm is the protein and the graph structure of input ligand $G(V,E)$, where $V$ contains one vertex for each of the rigid components and two components $(x,y)\in E$ iff rigid components $x$ and $y$ is connected by a single rotatable bond.

\begin{algorithm}
\caption{Anchor and grow subalgorithms of docking}
\SetAlgoLined
\DontPrintSemicolon
\SetKwFunction{anchor}{Anchor}
\SetKwFunction{grow}{Grow}
\SetKwFunction{growbond}{GrowBond}
\SetKwFunction{sort}{Sort}
\SetKwFunction{top}{Top}
\SetKwFunction{CompSize}{CompSize}
\SetKwFunction{GetPointCloud}{GetPointCloud}
\SetKwFunction{GetBestTransform}{GetBestTransform}
\SetKwFunction{Transform}{Transform}
\SetKwFunction{GravityCenter}{GravityCenter}
\SetKwFunction{LigandGen}{LigandGen}
\SetKwProg{Fn}{Function}{}{}
\BlankLine
\Fn{\anchor{$G(V,E), P$}}{
	$\boldsymbol{Q} \leftarrow \GetPointCloud{$P$}$\tcp*{Generate clustered point cloud for pocket in $P$}
	$\boldsymbol{M}' \leftarrow \emptyset$\;
	\ForEach{$x_1, x_2, x_3 \in V$ \tcp{triplet of connected rigid components}}{
		$\boldsymbol{M} \leftarrow \LigandGen{$\{x_1,x_2,x_3\}$}$\tcp*{\autoref{algoligprep}}
		\ForEach{$m \in \boldsymbol{M}$}{
			$t_1 \leftarrow \GravityCenter{$x_1$}$\;
			$t_2 \leftarrow \GravityCenter{$x_2$}$\;
			$t_3 \leftarrow \GravityCenter{$x_3$}$\;
			$\boldsymbol{T} \leftarrow \GetBestTransform{$\boldsymbol{Q},<t_1,t_2,t_3>$}$\;\tcp*{Get best transform matrix from triangle $<t_1,t_2,t_3>$ to points in $\boldsymbol{Q}$}
			$m' \leftarrow \Transform{$\boldsymbol{T},m$}$\;
			$\boldsymbol{M}' \leftarrow \boldsymbol{M}' \cup \{m'\}$\;
		}
	}
	$\boldsymbol{M}' \leftarrow \sort{$\boldsymbol{M}'$}$\tcp{by predicted energy potential}
	$\boldsymbol{M}' \leftarrow \top{$\boldsymbol{M}', K$}$\;
	\Return{$\boldsymbol{M}'$}\;
}
\BlankLine
\Fn{\grow{$G(V, E), P, \boldsymbol{M}$}}{
	$\boldsymbol{M}' \leftarrow \emptyset$\;
	\ForEach{$m \in \boldsymbol{M}$}{
		$\boldsymbol{R} \leftarrow \{(x,y) | x \in m, y \notin m, (x,y)\in E \}$\tcp*{Find growable bonds}
		\eIf{$\exists r_1, r_2 \in R, r_1, r_2$ is close} { \tcp{Grow single rotatable bond}
			$r \leftarrow \argmax_{(x,y)}{\CompSize{y},\forall (x,y)\in R}$\;
			\tcp*{Find the rotatable bond with the largest component}
			$\boldsymbol{M}' \leftarrow \boldsymbol{M}' \cup \growbond{m, r}$\;
		}{ \tcp{Independently grow all rotatable bonds}
			$\boldsymbol{U} \leftarrow \emptyset$\;
			\ForEach{$r \in \boldsymbol{R}$}{
				$\boldsymbol{C} \leftarrow \emptyset$\;
				\ForEach{$g \in \growbond{m, r}$}{
					$\boldsymbol{C} \leftarrow \boldsymbol{C} \cup \{ m\setminus g \}$\tcp*{Obtain grew part}
				}
				$\boldsymbol{U} \leftarrow \boldsymbol{U} \cup \boldsymbol{C}$\;
			}
			\ForEach{$t_1 \in \boldsymbol{U}_1, t_2 \in \boldsymbol{U}_2, ..., t_n \in \boldsymbol{U}_n$}{
				$m' \leftarrow m \cup t_1 \cup t_2 \cup ... \cup t_n $\;
				$\boldsymbol{M}' \leftarrow \boldsymbol{M}' \cup \{ m' \}$\;
			}
		}
	}
	\tcp{Merge nearby components}
	\ForEach{$m_1, m_2 \in \boldsymbol{M}$}{
		\If{$\exists x \in m_1, y \in m_2, m_1 \cap m_2 = \emptyset, (x, y) \in E$}{
			$m' \leftarrow m_1 \cup m_2$\;
		}
		$\boldsymbol{M}' \leftarrow \boldsymbol{M}' \cup \{ m' \}$\;
	}
	$\boldsymbol{M}' \leftarrow \sort{$\boldsymbol{M}'$}$\;
	$\boldsymbol{M}' \leftarrow \top{$\boldsymbol{M}', K$}$\;
	\Return{$\boldsymbol{M}'$}\;
}
\label{algoanchorgrow}
\end{algorithm}

\begin{algorithm}
\caption{Small compound docking algorithm}
\SetAlgoLined
\DontPrintSemicolon
\KwIn{Protein with target pocket $P$, graph structure of ligand to be docked $G(V,E)$}
\KwOut{a set of conformations, of the ligand in pocket $P$, sorted by predicted energy score, i.e. feasibility}
\SetKwFunction{anchor}{Anchor}
\SetKwFunction{grow}{Grow}
\SetKwFunction{dock}{Dock}
\SetKwFunction{VertexSet}{VertexSet}
\SetKwProg{Fn}{Function}{}{}
\Fn{\dock{Ligand $G(V,E)$, Protein $P$}}{
	$\boldsymbol{M} = $ \anchor{$G(V,E),P$}\;
	\While{$\VertexSet(\boldsymbol{M}) \neq V$}{
		$\boldsymbol{M} \leftarrow $ \grow{$G(V,E),P,\boldsymbol{M}$}\;
	}
	\Return{$\boldsymbol{M}$}\;
}
\label{algodock}
\end{algorithm}

\section{Experimental Results}
To show the actual performance of our model in real-life scenarios, several experiments have been set up, with comparisons with state-of-the-art methods in those particular fields.

\begin{comment}
\subsection{Ideal Conformation}
The first step of almost every molecular modeling task, be it conformation optimziation of small molecules or docking drug molecules into the target protein pocket, is to get good geometric constraints of molecules: that is, while the dihedral angles (around rotatable bonds) are relatively flexible in chemical and biochemical processes, other constrains such as bond lengths, bond angles, coplanarities are rather stable and thus compararable across different models. 

In practice, the Cambridge Structural Database(CSD)~\cite{CSD}, which stores over a million entries of X-Ray Diffraction data of a variety of molecules, are used as the ground truth for comparison.

Possible Benchmark Set:
\begin{itemize}
	\item  CSD: 800,896 entries with SMILES in our database: competent candidate: RDKit(MMFF,UFF), openbabel(MMFF,UFF), (possibly) OPENMM (used mainly for MD though)
	\item  Platinum Dataset~\cite{friedrich2017benchmarking} is used to compare different conformer generator.In another paper that demonstrates a fragment library based method for molecule generation~\cite{yoshikawa2019fast}, the \href{https://jcheminf.biomedcentral.com/articles/10.1186/s13321-019-0372-5/tables/1}{RMSDs} in bond/angle are given thus comparable.
	\item CSD x PubChem3D, as a direct comparison with OMEGA
	\item No executable of CORINA(which used a pseudo-force field for generation, presumably will not be very accurate), web interface might be used to run a reasonalbe size of dataset.
\end{itemize}
\end{comment}

\subsection{Side chain prediction}
In the publication that presented the program SCWRL4~\cite{krivov2009improved}, three kinds of descriptions of the accuracy of the model were given. Here the correctness is defined as having a difference of angle less than $40^{\circ}$, and the numbering of $\chi$ angles starts from the closest dihedral angle to the backbone:
\begin{enumerate}
\item The conditional probability that $\chi_i$ is correct, given $\chi_{i-1},\chi_{i-2},...\chi_1$ is correct, for all residues and for each type of amino acid.
\item The absolute probability that $\chi_{i},\chi_{i-1},...\chi_1$ is correct, for all residues and for each type of amino acid.
\item The root-mean-squared deviation (RMSD) of the side chain residues. The average RMSD value of a type of amino acid is calculated by averaging the sum of RMSD values of all residues of this type. The RMSD of a single residue is calculated using the formula:

\begin{equation}
RMSD\boldsymbol{(v,w)} = \sqrt{\frac{1}{n}\sum_{i=1}^n \norm{v_i-w_i}^2}
\end{equation}

\end{enumerate}

We have tested our model and SCWRL4 on SCWRL4's dataset comprising 379 PDB files. The results shown in \autoref{tableloocomp} show clear superiority of our method over SCWRL4 by having both lower RMSD and higher $\chi$ accuracies on all amino acids.

\begin{table} 
\centering
\caption{RMSD and $\chi$ value comparison between SCWRL4 and our method for LOO prediction on 379 proteins}
\begin{tabular}{c|c|cc|cc|cc|cc|cc|cc} 
\hline
\multirow{2}{*}{\begin{tabular}[c]{@{}c@{}}Amino\\Acid\end{tabular}} & \multirow{2}{*}{Count} & \multicolumn{2}{c|}{RMSD} & \multicolumn{2}{c|}{$\chi_1$} & \multicolumn{2}{c|}{$\chi_2$} & \multicolumn{2}{c|}{$\chi_3$} & \multicolumn{2}{c|}{$\chi_4$} & \multicolumn{2}{c}{$\chi_5$}  \\ 
\cline{3-14}
&          & Scwrl4 & Ours    & Scwrl4 & Ours & Scwrl4 & Ours & Scwrl4 & Ours & Scwrl4 & Ours & Scwrl4 & Ours \\ 
\hline
ALA & 5888 & 0.044  & 0.038   &      &        &      &        &      &        &      &        &      &        \\
ARG & 3719 & 1.866  & 1.365   & 83.0 & 88.9   & 71.6 & 78.6   & 48.4 & 61.4   & 38.0 & 52.1   & 38.0 & 52.1   \\
ASN & 2948 & 0.622  & 0.487   & 88.5 & 91.5   & 79.8 & 86.1   &      &        &      &        &      &        \\
ASP & 4142 & 0.621  & 0.458   & 88.0 & 92.4   & 79.6 & 86.0   &      &        &      &        &      &        \\
CYS & 1052 & 0.308  & 0.192   & 92.6 & 96.4   &      &        &      &        &      &        &      &        \\
GLN & 2590 & 1.157  & 0.919   & 82.8 & 87.6   & 67.2 & 74.9   & 52.0 & 66.5   &      &        &      &        \\
GLU & 4751 & 1.137  & 0.993   & 78.4 & 83.3   & 66.4 & 71.6   & 49.7 & 60.2   &      &        &      &        \\
GLY & 5547 & 0.000  & 0.000   &      &        &      &        &      &        &      &        &      &        \\
HIS & 1562 & 0.744  & 0.478   & 92.9 & 95.5   & 85.0 & 92.2   &      &        &      &        &      &        \\
ILE & 4058 & 0.352  & 0.281   & 96.6 & 97.6   & 86.7 & 90.0   &      &        &      &        &      &        \\
LEU & 6729 & 0.447  & 0.342   & 94.7 & 96.7   & 88.9 & 91.0   &      &        &      &        &      &        \\
LYS & 3995 & 1.414  & 1.216   & 80.1 & 87.0   & 68.6 & 76.0   & 54.1 & 59.3   & 35.4 & 38.4   &      &        \\
MET & 1430 & 0.970  & 0.651   & 86.6 & 92.8   & 75.9 & 87.3   & 59.9 & 71.7   &      &        &      &        \\
PHE & 2800 & 0.576  & 0.304   & 97.4 & 99.3   & 95.9 & 98.9   &      &        &      &        &      &        \\
PRO & 3319 & 0.210  & 0.173   & 87.0 & 91.6   & 83.3 & 88.0   &      &        &      &        &      &        \\
SER & 4210 & 0.539  & 0.416   & 72.8 & 80.4   &      &        &      &        &      &        &      &        \\
THR & 3920 & 0.316  & 0.247   & 90.8 & 93.8   &      &        &      &        &      &        &      &        \\
TRP & 1008 & 1.001  & 0.432   & 94.7 & 98.7   & 87.7 & 96.0   &      &        &      &        &      &        \\
TYR & 2416 & 0.687  & 0.379   & 97.2 & 99.2   & 95.4 & 98.4   &      &        &      &        &      &        \\
VAL & 5138 & 0.261  & 0.211   & 93.4 & 95.3   &      &        &      &        &      &        &      &        \\
\hline
\end{tabular}
\label{tableloocomp}
\end{table}

\subsection{Docking}

The testing set is our filtered protein data bank (PDB) database for the purpose of reliable benchmarking: PDB Docking Set v2 (PDSv2). 

The criteria are listed as follow:
\begin{enumerate}
\item The PDB structure is determined by X-ray diffraction with a resolution $<2.5\angstrom$
\item The ligand should belong to a protein with type "protein".
\item The ligand should be connected (in a graph theory sense), NOT being a solvent molecule, and not bonded to any atom that is not part of itself.
\item The ligand should have $>5$ non-hydrogen atoms in the ligand, no more than 8 atoms in its largest simple ring, and $\le10$ rotatable bonds.
\item There should be no external metal atoms within $3.0\angstrom$ from the ligand. 
\item There should be no hetero atom with the same Residue name within $5.0\angstrom$
\item There should be external atoms within $3.0\angstrom$ from the metal atom in the ligand
\item The ratio of complete amino acids, i.e., amino acids without missing atoms, is larger than $90\%$
\end{enumerate}

A total of 1441 high quality structures of protein-ligand complex are selected this way for benchmarking.

We do our testing against several other widely-accepted docking algorithms, including UCSF Dock~\cite{allen2015dock}, AutoDock Vina~\cite{trott2010autodock}, and Rosetta~\cite{deluca2015fully}. The results are shown in \autoref{tabdockcomp}. RMSD is calculated between positions of each docked atom $v_i$ and its corresponding ground truth position $w_i$. Special treatment is done to handle symmetrical positions so that flipping a benzene ring will result in zero RMSD. We also considered a modified RMSD metric called the shape RMSD, wherein for each docked atom $v_i$, the position $w_i$ corresponds to the nearest atom in the ground truth conformation, which may not be the same atom in the molecule. A low shape RMSD indicates the algorithm is able to fit the molecule into the shape of the pocket.

It is clear from the results that our machine-learned model performs much better than all previous state-of-the-art methods.

\begin{table}
\caption{RMSD and shape RMSD comparison between our methods and other methods. ($\text{@}k$ indicates the best evaluation score of the first $k$ of the predicted conformations, sorted by the algorithm's internal ranking score.)}
\begin{tabular}{l|c|c|c|c|c}
\hline
                   & ShapeRMSD@1 & ShapeRMSD@5 & RMSD@1      & RMSD@5      & Success Count \\
\hline
Ours               & 0.858651293 & 0.637538778 & 1.769472363 & 1.032891272 & 1441          \\
UCSF-FLX-FLEXIBLE  & 1.352609299 & 0.890155312 & 2.634079219 & 1.484691792 & 1431          \\
UCSF-FLX-RIGID     & 1.831002016 & 1.250156037 & 3.814842201 & 2.497495059 & 1433          \\
AutoDock Vina      & 1.354831147 & 0.895898547 & 2.770003983 & 1.586554709 & 1433          \\
Rosetta            & 1.291723616 & 0.923275125 & 2.755971552 & 1.728418176 & 1418          \\
\hline
\end{tabular}
\label{tabdockcomp}
\end{table}

\section{Related work} \label{secrelatedwork}

\subsection{Force field for molecule optimization} \label{subsubsecmolconfopt}

Once a smooth, accurate, differentiable energy function of atom positions has been obtained, the most common use of such a function is conformation optimization. Most molecule modeling tasks involve conformation optimization, i.e., searching for a low-energy, stable, thus most probable 3D position configuration of atoms. While higher-level quantum chemistry theory exists, the computational complexity becomes prohibitive for most tasks, e.g., large biomolecules, virtual screening of millions of molecules, etc. So, in the development of empirical force fields, compromises must be made, usually in favor of fitting some specific system's properties.

To achieve this, the force field first needs to assign proper constraints of bond lengths and angles of the molecules, and then sample and optimize the most probable conformations. The process is usually referred to as the "ligand preparation step". Prior major force fields that have been developed to solve the molecular modeling task include:

\begin{itemize}
	\item OMEGA~\cite{hawkins2010conformer}, which has been chosen as the 3D model generator for one of the largest online compound database PubChem~\cite{pubchem3d}.
	\item Merk Molecular Force Field (MMFF)~\cite{halgren1996merck} and its variants, which have been developed by Merck and are mainly used in the field of small drug-like molecules. 
	\item CHARMM General Force Field (CGENFF)~\cite{vanommeslaeghe2012automationcharmm}, which is part of the biomolecular force field CHARMM and designed specifically for small organic molecules. 
	\item Universal Force Field (UFF)~\cite{rappe1992uff}, which is the only force field that claims to include every element in the periodic table. It features a small number of parameters, which can be fit in an A4 paper.
	\item CORINA~\cite{gasteiger1990automatic}, which is mainly used for ligand conformation generation.
\end{itemize}

Other popular force fields include AMBER~\cite{wang2004development}, OPLS~\cite{damm1997opls}, and GROMOS~\cite{schmid2011definition}. Most of these force fields are designed and parameterized for specific systems, which means some have particular strengths in some systems while having weaknesses in others. A general force field fitting all particular systems, such as organic chemicals, biochemical compounds, and proteins, is yet to be developed.

\subsection{Side-chain conformation prediction}
Predicting side-chain conformations, given a rather fixed backbone, is a crucial part of many protein-related tasks, such as docking and homology modeling, wherein the conformational changes of one or several side chains in the pocket or of the mutated amino acids are pivotal to solving the problem. 

SCWRL4~\cite{krivov2009improved} uses a backbone-dependent rotamer library for discrete sampling. It combines the CHARMM force field with some hand-tuned specialized potential terms as the energy function. The main focus is a sophisticated method for predicting all side chains of a protein.

In our previous work~\cite{sidechainpaper} we show that side chain prediction can be greatly improved with neural networks. We can transform a particular side chain into a 3D grid and then train a 3D convolutional neural network to predict the fitness score. Despite the improved performance, the lack of interpretability of this black box model still persists and undermines further implementation into other modeling tasks, compared to explicit force fields. 

\subsection{Ligand docking}
There are many small molecule docking programs currently available. Most are based on a sampling-then-scoring methodology as they combine a custom sampling algorithm with a traditional molecular force field. For example, UCSF Dock~\cite{allen2015dock} uses an anchor and grow strategy for position sampling. Whereas AutoDock Vina~\cite{trott2010autodock} focuses on the scoring function optimization and uses an Iterated Local Search global optimizer with a BFGS local optimizer for conformation optimization. Rosetta~\cite{deluca2015fully} uses a multi-scale Monte Carlo-based method for perturbating and sampling the possible ligand conformations. Finally, GLIDE~\cite{halgren2004glide} uses pruned exhaustive search for the initial phase and Monte Carlo-based method for the grow phase.

\section{Conclusion} \label{secconclusion}

Molecular force field construction has many different applications in drug discovery and molecular modeling. In this paper, we show a novel method of using neural networks to train a potential function. It combines the benefits of traditional handcrafted potential functions as being smoothly differentiable, with the benefits of being fully automatically-trained from large crystal structure databases. We tested the trained potential function and showed it has superior performance over existing molecular force fields, without the need of any manual parameter tuning.

\bibliographystyle{unsrt}
\bibliography{references}  %%% Uncomment this line and comment out the ``thebibliography'' section below to use the external .bib file (using bibtex) .

\end{document}